 \documentclass[final,5p,times,twocolumn,authoryear]{elsarticle}

\usepackage{amssymb}
\usepackage{lipsum}

\usepackage{amsmath} 

\usepackage{algorithm}
\usepackage{algorithmic}
\usepackage{booktabs}
\usepackage{multirow} 
\usepackage{colortbl}
\usepackage{xcolor}

\journal{Results in Physics}

\begin{document}

\begin{frontmatter}



\title{Design-MLLM: A Reinforcement Alignment Framework for \\Verifiable and Aesthetic Interior Design}


\author[a]{Yuxuan Yang}
\ead{qingzhuo13@njfu.edu.cn}
\author[d,e]{Xiaotong Mao}
\ead{mao12u@etu.univ-lorraine.fr}
\author[b,c]{Jingyao Wang}
\ead{wangjingyao2023@iscas.ac.cn}

\affiliation[a]{organization={Nanjing Forestry University},
            city={Nanjing},
            country={China}}

\affiliation[b]{organization={Institute of Software Chinese Academy of Sciences},
            city={Beijing},
            country={China}}

\affiliation[c]{organization={University of Chinese Academy of Sciences},
            city={Beijing},
            country={China}}

\affiliation[d]{organization={Université de Lorraine},
            city={Nancy},
            country={France}}

\affiliation[e]{organization={Laboratoire Réactions et Génie des Procédés},
            city={Nancy},
            country={France}}

\begin{abstract}
Interior design is a requirements-to-visual-plan generation process that must simultaneously satisfy verifiable spatial feasibility and comparative aesthetic preferences. While recent multimodal large language models (MLLMs) offer a unified foundation for interpreting user intent and producing design rationales, our empirical analysis reveals a persistent contradiction in real-world deployment: MLLMs often produce layouts that are unbuildable and aesthetically inconsistent. These findings indicate that simply adding in-domain text is insufficient; effective interior design requires an alignment mechanism that separates hard constraints from soft preferences and coordinates them during optimization.
To address this, we propose Design-MLLM, a reinforcement alignment framework that optimizes a feasibility-first preference objective via a dual-branch, aesthetic-oriented reward. Specifically, Design-MLLM (i) explicitly evaluates spatial feasibility using programmatic constraint checks, (ii) assesses aesthetic preference only among feasible candidates to avoid visually appealing but unexecutable shortcuts, and (iii) performs group-relative optimization to obtain stable preference signals. 
Through this process, Design-MLLM learns a controllable policy that consistently selects and generates solutions that are both executable and aesthetically coherent, rather than occasionally producing visually appealing but infeasible designs.
Extensive experiments on various benchmark datasets demonstrate the advantages of Design-MLLM.
\end{abstract}




\end{frontmatter}




\section{Introduction}
\label{sec:intro}
Interior design \cite{ching2018interior,ulrich1991effects,pile2005history} is fundamentally a generative process that translates functional needs, aesthetic preferences, and geometric constraints into deliverable visual plans such as floor layouts and renderings. This process necessitates verifiable structural decisions regarding scale, ergonomics, circulation, and functional zoning, while simultaneously satisfying aesthetic expectations regarding style, color, materials, lighting, and atmosphere. With the advancement of machine learning, numerous techniques have been proposed to assist distinct stages of this pipeline, including concept sketch generation \cite{shah2001collaborative,yang2003concept}, rendering synthesis \cite{hu2020graph2plan,geng2025diffdesign}, style recommendation \cite{de2021style,chao2009framework}, and material coordination \cite{zhang2023adding}. These innovations have significantly reduced iteration costs and improved overall design efficiency.

Recent progress in multimodal large language models (MLLMs) further creates opportunities for automating interior design \cite{liu2023visual,ccelen2024design,gallega2025exploring,guo2025copo}. Trained on large-scale data, MLLMs can unify multimodal information for reasoning, making them promising foundations for design agents \cite{chen2024mllm,li2024surveying,kuang2025natural}: they can interpret user requirements, parse spatial semantics, and provide clear language explanations that guide downstream generators, such as visual drafts or structured layouts. However, generic MLLMs are primarily pre-trained on open-domain web corpora for description, question answering, and recognition, and thus lack systematic modeling of the coupled relationships among scale, layout, function, and style that are essential in interior design \cite{kim2025autopaperbench,wang2024genartist,wang2025chat2layout,wang2024roomdreaming,feng2025follow}. To adapt MLLMs to this domain, concurrent efforts mainly fall into two categories \cite{lai2024llms,wang2025omnigenbench,chen2025mindgpt}: (i) instruction-translation approaches that rewrite user requirements into prompts for rendering models or fixed pipelines, and (ii) heavily supervised approaches that rely on expensive expert annotations (e.g., layouts, rules, and style labels). Despite partial progress, they remain limited by insufficient verifiable spatial reasoning and high training cost \cite{zhong2024topv,wu2025spatial,zheng2025multimodal}. Therefore, this paper investigates two questions: (i) what fundamentally hinders the application of MLLMs to interior design, and (ii) how to overcome these obstacles.

Unlike generic image generation, interior design requires jointly satisfying spatial feasibility and aesthetic preference. The spatial side corresponds to a programmatically verifiable feasible set. For example, out-of-boundary and collision avoidance, door clearance, passage width, circulation connectivity, and window occlusion; whereas the aesthetic side is inherently comparative, involving style consistency, color psychology, material logic, density-whitespace balance, and atmosphere narrative. Consequently, interior design generation is not merely ``producing attractive images'', but a reasoning and decision-making problem of selecting preferred solutions within a feasible set. In real-world settings, existing models repeatedly expose a core contradiction: hard-constraint usability and soft-preference aesthetics are difficult to satisfy simultaneously and consistently. For example, models may produce plausible descriptions or visually appealing solutions, yet the resulting layouts can block doors, collide, exhibit distorted proportions, or show stylistic collage and drift, making them unsuitable for deliverable design. This indicates that current paradigms struggle to jointly optimize feasibility and aesthetics, which substantially limits the practical deployment of MLLMs in interior design.
Our experiments further validate this conclusion (\textbf{Subsection \ref{sec:emprical}} and \textbf{Figure \ref{fig:motivation_stats}}). Under two types of test scenarios, we observe three stable failure modes: (i) insufficient spatial feasibility: semantic recognition does not translate into geometric executability, while the model can identify furniture and suggest placements, its structured layouts often violate constraints such as out-of-boundary errors, object overlap, and door clearance, yielding outputs that are describable but not buildable; (ii) unstable aesthetic consistency: although the model can articulate style and ambience, it weakly constrains color relations, material coordination, and style coherence, leading to collage-like conflicts or multi-round drift that undermines overall unity and professional credibility; and (iii) rationale-output decoupling: the model’s rationales frequently fail to constrain layout fields and material specifications, causing explanation-output inconsistency and amplifying hallucination risk. \textbf{These results suggest that interior design cannot be solved by simply adding in-domain text; rather, it requires an alignment mechanism that explicitly separates and coordinates hard constraints and soft preferences at the training-objective level, so the model learns to become ``more aesthetic within the feasible domain'', instead of merely becoming more fluent in language.}

Based on the above analyses, we propose Design-MLLM, a reinforcement alignment framework for interior design. The core idea is to align generation with a feasibility-first preference objective through a dual-branch aesthetic-oriented reward: (i) a spatial feasibility branch, which explicitly measures whether a candidate design satisfies verifiable constraints such as collisions, clearances, reachability, and circulation; and (ii) an aesthetic preference branch, which evaluates consistency with user intent in style, color, material, and ambience only among feasible candidates, preventing the model from exploiting ``visually appealing but unbuildable'' shortcuts. Taken together, Design-MLLM shifts supervision from producing occasional good-looking images to learning a controllable policy that consistently generates designs that are both executable and aesthetically coherent.
Specifically, Design-MLLM consists of three steps: (i) Feasibility-guided candidate generation: for each user request, the model produces a group of candidate solutions and automatically verifies their geometric and functional validity via constraint checks; (ii) Decoupled reward construction: we compute spatial feasibility feedback and aesthetic preference feedback separately, and perform relative comparisons within the feasible set to obtain stable preference signals; (iii) policy optimization: we convert these groupwise signals into fine-grained learning targets and update the model with a GRPO-style objective, encouraging the policy to prefer solutions that are more aesthetic within the feasible region. Extensive experiments on various benchmark datasets demonstrate that Design-MLLM consistently improves both spatial executability and aesthetic alignment across diverse interior design scenarios.

The main contributions can be summarized as:
\begin{itemize}
    \item We provide a systematic empirical diagnosis of why generic MLLMs fail in deployable interior design, and identify a persistent feasibility, i.e., aesthetics conflict. The models may generate fluent rationales or visually appealing renderings, yet frequently violate programmatically verifiable constraints and exhibit unstable aesthetic consistency and rationale-output mismatch.
    \item We propose Design-MLLM, a reinforcement learning post-training framework that decouples hard constraints from soft preferences through a dual-branch reward: a spatial-feasibility branch for explicit constraint verification and an aesthetic-preference branch that performs relative comparison only within the feasible set, optimized with a GRPO-style group-relative objective to learn more aesthetic within the feasible domain.
    \item Extensive experiments across multiple benchmark datasets demonstrate that Design-MLLM consistently improves both spatial executability and aesthetic alignment. Comprehensive ablations further verify the necessity of each component of our method and why it performs well.
\end{itemize}

\section{Related Work}

\subsection{Interior Design}

Interior design has long been studied as a structured generation and decision-making problem that must satisfy both functional requirements and spatial constraints, while producing visually coherent outcomes. 
Early computational approaches for interior design focused on representing rooms, objects, and their relations with structured abstractions, then searching for layouts that satisfy geometric and functional rules. Recent learning based methods shift this pipeline toward data-driven generation, where the model learns priors over spatial organization and style regularities from large collections of designed spaces. These works commonly treat interior design as a structured generation problem that must preserve functional zoning, circulation, ergonomics, and object-level feasibility \cite{ching2018interior,pile2005history,ulrich1991effects}.

A major line of work studies automatic floorplan and layout synthesis under explicit structural constraints. House GAN models layout generation conditioned on graph constraints that encode room types and adjacencies, enabling graph-constrained generation of room boxes \cite{nauata2020house}. Graph2Plan conditions on a layout graph and a building boundary, and learns to generate floorplans that satisfy the boundary and room relationship constraints \cite{hu2020graph2plan}. Beyond architectural floorplans, LayoutGAN and related methods model geometric relations among elements for layout realism, which provides useful abstractions for furniture placement and arrangement in interior scenes \cite{li2019layoutgan}. 
Another line of work targets indoor scene synthesis with explicit object-level structure and datasets. PlanIT introduces a plan and instantiate framework that first generates a relation graph and then instantiates concrete object placements guided by spatial priors \cite{wang2019planit}. ATISS formulates indoor scene synthesis with autoregressive transformers over unordered object sets, improving controllability and realism for furniture layouts under a given floor plan \cite{paschalidou2021atiss}.
In parallel, rendering synthesis and controllable visualization have become central to modern design workflows, as they enable rapid exploration of style, materials, and ambience. Diffusion models and latent diffusion frameworks have substantially improved the fidelity of image synthesis~\cite{rombach2022high}, and controllable generation techniques such as ControlNet allow structure-guided rendering from cues like sketches, edges, depth, or segmentation~\cite{zhang2023adding}. 
These advances significantly reduce the cost of visual iteration. 

However, they primarily optimize for perceptual realism and conditioning adherence, leaving a gap between visually attractive renderings and deliverable, constraint-satisfying design plans. More specifically, structurally valid layouts may lack aesthetic nuance, while visually stunning renderings frequently hallucinate physically impossible geometries. This gap becomes particularly evident in interior design, where small geometric violations can invalidate an otherwise plausible proposal.
To overcome this, we propose Design-MLLM, which bridges this gap by integrating both spatial and aesthetic reasoning directly into the generative policy of an MLLM. Instead of treating feasibility constraints and aesthetic styles as separate pipeline steps, we model them as a unified decision-making process, ensuring that the generated designs retain the structural rigor of rule-based systems while benefiting from the semantic richness of modern generative models.

\subsection{Multimodal Large Language Models}
Multimodal large language models (MLLMs) connect a visual encoder with a large language model to support joint reasoning over images and text. Flamingo \cite{alayrac2022flamingo} demonstrates strong few-shot vision language generalization via gated cross attention, establishing an effective connector design between modalities. BLIP \cite{li2023blip} improves training efficiency by freezing major backbones and learning a lightweight querying module to bridge vision and language, which makes it attractive for domain adaptation where full end-to-end retraining is costly. Instruction-tuned multimodal assistants such as LLaVA \cite{liu2023visual} extend language instruction tuning to the vision language setting, improving interactive reasoning and following user intent.

The alignment methods are critical when MLLMs are used as decision-making policies rather than pure predictors. InstructGPT \cite{ouyang2022training} popularizes reinforcement learning from human feedback to align outputs with user preferences, typically using a reward model and policy optimization. PPO \cite{schulman2017proximal} is a standard policy gradient backbone used in many RLHF-style pipelines due to its stability under clipped objectives. Direct preference optimization \cite{rafailov2023direct} simplifies preference alignment by optimizing a closed-form objective derived from pairwise preferences, reducing the need for explicit reward modeling in many settings. Recently, group relative policy optimization (GRPO) \cite{shao2024deepseekmath,gu2025group} that normalizes rewards within sampled candidate groups further improves training efficiency and stability for autoregressive models, and have been adopted in recent reasoning-focused alignment systems.

For interior design generation, the alignment advances enable models to parse user requirements, reason about spatial semantics, and produce structured plans or renderings. However, generic alignment signals are often insufficient because interior design requires simultaneous satisfaction of verifiable hard constraints and comparative soft preferences \cite{ching2018interior,ulrich1991effects,pile2005history}. This gap motivates alignment mechanisms that explicitly separate feasibility verification from aesthetic preference learning and that update the policy toward being more aesthetic within the feasible region rather than merely improving linguistic fluency.

\section{Problem Settings and Analyses}

\subsection{Problem Settings}
\label{sec:problem_settings}

\paragraph{\textbf{Multimodal Large Language Models (MLLMs)}} Consider a paired image-text dataset
$\mathcal{D}=\{(I_v^{(n)}, I_t^{(n)}, Y^{(n)})\}_{n=1}^{N}$,
where $I^{(n)}=(I_v^{(n)}, I_t^{(n)})$ comprises an image $I_v^{(n)}$ and a text prompt $I_t^{(n)}$, and $Y^{(n)}$ denotes the ground-truth target.
An MLLM first maps the image $I_v$ into a sequence of visual tokens using a vision encoder (e.g., ViT), and tokenizes the prompt $I_t$ with a language tokenizer.
The resulting visual and textual tokens are concatenated and fed into the language model, which autoregressively generates an output token sequence
$o=\{o_1,o_2,\ldots,o_T\}$ with:
\begin{equation}
    p_\theta(o \mid I) = \prod_{t=1}^{T} p_\theta(o_t \mid I, o_{<t}).
\end{equation}
where $T$ is the sequence length and $o$ may include both intermediate tokens (e.g., reasoning traces) and final answer tokens.

Equivalently, letting $H$ denote the fused multimodal representation produced by the model, the autoregressive policy can be written as:
\begin{equation}
    p_\theta(y\mid H)=\prod_{t=1}^{T}p_\theta(y_t\mid H,y_{<t})=\prod_{t=1}^{T}g_\theta(h_t,y_{<t}),
\end{equation}
where $g_{\theta}(\cdot)$ returns the next-token distribution conditioned on $H$ and the previously generated tokens.

To achieve precise inference, the MLLM is typically trained by maximizing the average log-likelihood over $\mathcal{D}$:
\begin{equation}
    \max_{\theta} \frac{1}{N} \sum_{n=1}^N \sum_{t=1}^{T^{(n)}} \log p_\theta(o_t^{(n)} \mid I^{(n)}, o_{<t}^{(n)}).
\end{equation}

Taking the widely used reinforcement learning-based post-training method as an example, we can optimize the model parameters $\theta$ by maximizing the reward under the autoregressive output policy $\pi_\theta(\tilde{Y}|I)=p_\theta(\tilde{Y}|I)$. The objective can be expressed as:
\begin{equation}
    \max_\theta \, \mathbb{E}_{I\sim\mathcal{D}}\, \mathbb{E}_{\tilde{Y}\sim\pi_\theta(\cdot\mid I)}[r(\tilde{Y})]
\end{equation}
where $r(\cdot)$ denotes a task-specific reward (e.g., BLEU~\cite{papineni2002bleu} or ROUGE~\cite{lin2004rouge}) that compares the model output $\tilde{Y}$ with the ground truth $Y$.

For inference, given a new input $I=(I_v, I_t)$, the model decodes a token sequence according to $p_{\theta}(o\mid I)$ (e.g., via greedy or beam search) and extracts the final answer $\tilde{Y}$ from $o$.
Let $\mathcal{O}$ denote the space of semantic objects relevant to the input (e.g., entities, attributes, and events), and let $\mathcal{O}_{\text{true}}\subseteq \mathcal{O}$ be the set of ground-truth semantic targets present in the multimodal input.
From the generated sequence $o=\{o_1,\ldots,o_T\}$, we extract the mentioned objects $\mathcal{O}_{\text{gen}}\subseteq \mathcal{O}$ for downstream evaluation.

Beyond discriminative tasks, MLLMs can be used as conditional generators that translate multimodal user intent into structured or visual outputs.
In interior design generation, the conditioning signal typically includes textual requirements (e.g., functional needs, style preferences, budget) and optional visual references (e.g., floorplans, sketches, exemplars).
Formally, we write the conditioning input as
$X=(I_v, I_t, C)$, where $C$ denotes additional structured constraints (e.g., room sizes, door/window locations, adjacency rules).
The model then generates a design output sequence $o\sim p_\theta(\cdot\mid X)$, from which we parse a design proposal $\hat{S}=\phi(o)$ (e.g., layout plan, object placements, material/color specifications, or render prompts).
This process can be summarized as
\begin{equation}
\hat{S}=\phi(o),\quad o \sim p_\theta(o\mid X),\quad
p_\theta(o\mid X)=\prod_{t=1}^{T}p_\theta(o_t\mid X,o_{<t}).
\end{equation}
Moreover, interior design is inherently constraint-heavy: for each input $X$, hard spatial rules induce a feasible set $\mathcal{F}(X)$ of executable designs.
An ideal generator should concentrate probability mass on feasible and preference-aligned solutions, i.e.,
$p_\theta(\hat{S}\in \mathcal{F}(X)\mid X)$ is high, while among $\mathcal{F}(X)$ it further favors designs that match the user's aesthetic preferences.

\paragraph{\textbf{A Brief Introduction to GRPO for MLLMs}} 
Given a paired image-text dataset
$\mathcal{D}=\{(I_v^{(n)}, I_t^{(n)}, Y^{(n)})\}_{n=1}^{N}$,
each sample consists of an input $I^{(n)}=(I_v^{(n)}, I_t^{(n)})$, i.e., an image $I_v^{(n)}$ and a text prompt $I_t^{(n)}$, together with its ground-truth answer $Y^{(n)}$.
For each input $I^{(n)}$, GRPO draws a group of $G$ candidate sequences $\{o^{i}\}_{i=1}^{G}$ from the old policy $\pi_{\theta_{\rm old}}(\cdot \mid I^{(n)})$.
Each candidate $o^{i}=\{o^{i}_{1},\ldots,o^{i}_{T_i}\}$ contains both intermediate tokens (e.g., reasoning steps) and answer tokens.
We denote the extracted answer portion by
$\tilde{Y}^{i}=\{\tilde{y}^{i}_{1},\ldots,\tilde{y}^{i}_{T_{y^{i}}}\}$,
and regard the remaining tokens $o^{i}\setminus \tilde{Y}^{i}$ as intermediate reasoning.
The updated policy $\pi_{\theta}$ is then learned by maximizing the following objective:
\begin{equation}\label{eq:grpo_mllm}
\begin{aligned}
\mathcal{J}_{\rm GRPO}
= \mathbb{E}&_{[I\sim P,\ \{o^i\}\sim \pi_{\theta_{\rm old}}]} \\
\frac{1}{G}\sum_{i=1}^{G}\frac{1}{T_i}\sum_{t=1}^{T_i}
 \min(R_{i,t}(\theta)&A_i,\ \Psi(A_{i,t}))
-\beta\; \mu(\pi_\theta)),\\
\end{aligned}
\end{equation}
where $\epsilon$ and $\beta$ are hyperparameters, and $\Psi(A_{i,t})$ denotes the clip operator, i.e., $\Psi(A_{i,t})=\mathrm{clip}\!\left(R_{i,t},\,1-\epsilon,\,1+\epsilon\right)\cdot A_{i,t}$.
We regularize the update with $\mu(\pi_\theta)=D_{\mathrm{KL}}(\pi_\theta \Vert \pi_{\mathrm{ref}})$, and $T_i$ denotes the length of the sampled sequence
$o^i=\{o^i_1,\ldots,o^i_{T_i}\}$.
The group-relative advantage $A_i$ is computed within each candidate set to reflect the comparative quality of outputs:
\begin{equation}\label{eq:adv}
A_i=\frac{r^i-\mathrm{mean}(r^1,\ldots,r^{G})}{\mathrm{std}(r^1,\ldots,r^{G})},
\end{equation}
where $r^{i}=\mathrm{reward}(Y^{i})$ denotes the outcome reward~\cite{guo2025deepseek}, computed from the correctness of the extracted answer $\tilde{Y}^{i}$ (e.g., $r^{i}=1$ if $\tilde{Y}^{i}$ is correct and $r^{i}=0$ otherwise).
Here, $\mathrm{mean}(\cdot)$ and $\mathrm{std}(\cdot)$ are the mean and standard deviation evaluated over the $G$ rewards in the same group.
The token-level importance ratio can be expressed as:
\begin{equation}\label{eq:ratio}
R_{i,j}(\theta)=\frac{\pi_\theta(o^{i}_t \mid I^{(n)},\ o^{i}_{<t})}{\pi_{\theta_{\rm old}}(o^{i}_t \mid I^{(n)},\ o^{i}_{<t})}.
\end{equation}
The KL regularizer is evaluated at the token level against a fixed reference policy $\pi_{\rm ref}$ (typically chosen as $\pi_{\theta_{\rm old}}$) as:
\begin{equation}\label{eq:kl}
D_{\rm KL}(\pi_\theta\Vert \pi_{\rm ref})=\frac{\pi_{\rm ref}(o^{i}_t | I^{(n)},\ o^{i}_{<t}\big)}{\pi_\theta(o^{i}_t | I^{(n)},\ o^{i}_{<t})}-\log\frac{\pi_{\rm ref}(o^{i}_t | I^{(n)},\ o^{i}_{<t})}{\pi_\theta(o^{i}_t | I^{(n)},\ o^{i}_{<t})}-1.
\end{equation}

\begin{figure*}[t]
    \centering
    \includegraphics[width=\linewidth]{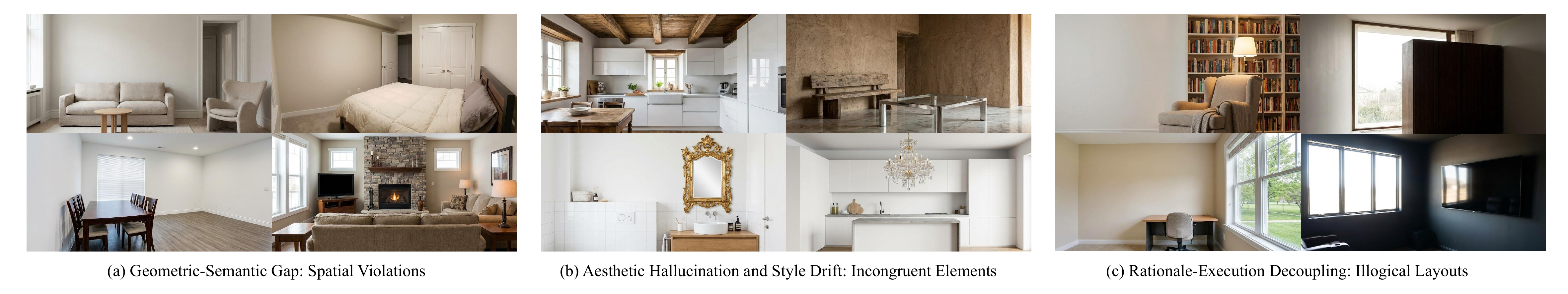}
    \caption{Examples of critical deficiencies identified in our empirical analysis.}
    \label{fig:motivation_vis}
\end{figure*}

\begin{figure*}[t]
    \centering
    \includegraphics[width=\linewidth]{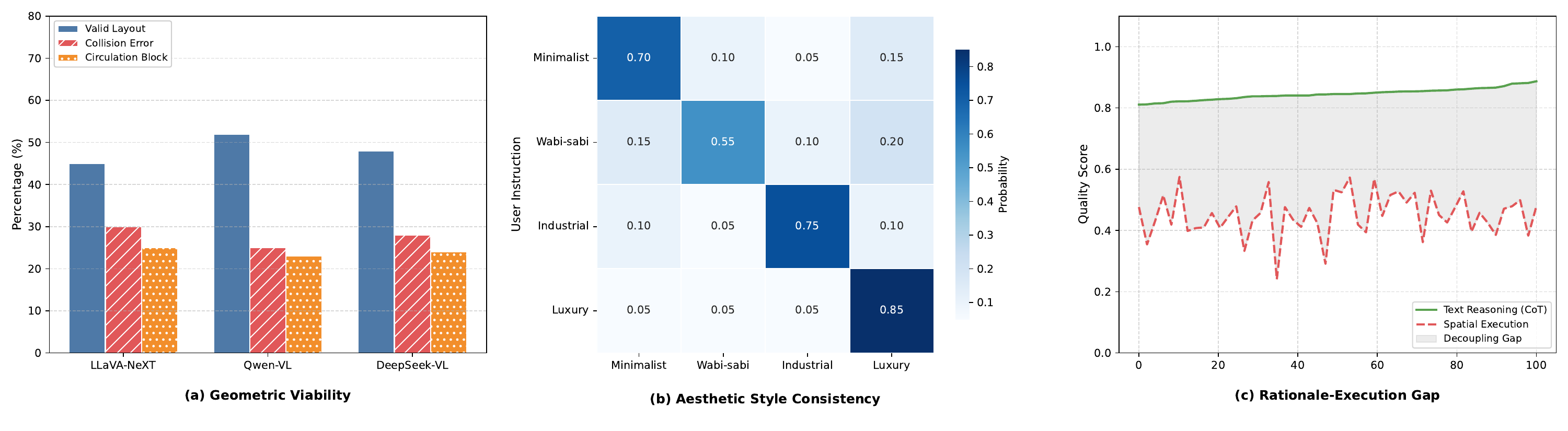}
    \caption{Empirical Analysis.
    (a) Geometric Viability: Models frequently violate hard constraints like collisions and circulation blockage. 
    (b) Style Drift: Confusion matrix showing the probability of models hallucinating ``Modern Luxury'' elements when prompted for ``Wabi-sabi'' or ``Minimalist'' styles due to data bias. 
    (c) Rationale-Execution Gap: A persistent decoupling where high-quality textual reasoning (CoT) does not translate into high-quality spatial execution.}
    \label{fig:motivation_stats}
\end{figure*}

\subsection{Empirical Analyses}
\label{sec:emprical}
To diagnose the failure modes of current MLLMs in interior design, we conduct a series of experiments for evaluation.

Specifically, we employ representative open-source MLLMs, including LLaVA-NeXT, Qwen-VL, and DeepSeek-VL, to act as design reasoning agents. Their textual rationales and structured layout suggestions are subsequently used to condition advanced generative models (e.g., Stable Diffusion XL \cite{podell2023sdxl} or Midjourney) to produce final visual designs for evaluation. We design a Design-QA task and a Layout-Generation task, where we input floor plan images alongside textual requirements (e.g., ``A minimalist living room with warm lighting and high traffic flow efficiency''). Some generated results are shown in \textbf{Figure \ref{fig:motivation_vis}}.

We evaluate the outputs of existing methods against professional standards derived from architectural handbooks \cite{pickard2008architects}. The analysis reveals three critical deficiencies, as illustrated in \textbf{Figure \ref{fig:motivation_stats}}:
(i) The Geometric-Semantic Gap (The Scale Problem). While open-source MLLMs excel at identifying semantic objects (e.g., sofa, window), they exhibit severe geometric blindness. In approximately 40\% of generated structured layouts, we observe anthropometric violations. For instance, models frequently place coffee tables within 300mm of sofas (violating the standard 400-450mm legroom requirement) or position furniture in a way that obstructs the circulation path (the primary movement artery between zones). This indicates that these models perceive design as a ``bag-of-words'' visual composition rather than a scale-sensitive spatial arrangement. (ii) Aesthetic Hallucination and Style Drift (The Harmony Problem). Aesthetic consistency requires maintaining a unified visual language across geometry, texture, and color. However, current open-source models suffer from significant style drift. When prompted for specific styles like Wabi-sabi (characterized by asymmetry, roughness, and natural materials), the models often introduce incongruent elements such as Modern Luxury textures (e.g., high-gloss marble or chrome) simply because these tokens co-occur frequently in web-scale training data. Furthermore, they struggle with chromatic harmony. Although the models describe color palettes correctly (e.g., earthy tones), the generated visual plans often violate color theory principles, such as failing to balance dominant, sub-dominant, and accent colors (the 60-30-10 rule), resulting in visual noise rather than atmospheric coherence.
(iii) Rationale-Execution Decoupling. A profound issue lies in the misalignment between the model's reasoning chain and its final output. We observe that models correctly reason about ``maximizing natural light'' in their textual response, yet paradoxically place tall storage units directly in front of south-facing windows in the generated layout. This suggests that the model's aesthetic reasoning is purely linguistic mimicry, detached from its spatial decision-making modules.

\subsection{Motivation Analyses}
\label{sec:motivation}
The empirical deficiencies observed in Section \ref{sec:emprical} represent more than mere performance fluctuations; they expose a fundamental misalignment between the probabilistic nature of generic MLLMs and the rigorous constraints of interior design. We analyze this misalignment from the perspectives of training objectives and optimization landscapes to elucidate why a tailored alignment framework is indispensable.

\emph{Probabilistic Plausibility vs. Physical Validity.} The core conflict stems from the objective function of generic MLLMs. These models are trained to maximize the likelihood of the next token based on large-scale web corpora, essentially optimizing for \textit{semantic plausibility}, what looks or sounds probable based on data distribution. However, interior design demands \textit{physical validity}, which is governed by strict, non-negotiable rules of geometry and ergonomics. In the training data, physical constraints (e.g., a chair cannot overlap with a wall) are implicit latent variables rather than explicit supervision signals. Consequently, MLLMs learn to mimic the visual surface of design renders without comprehending the underlying constructive logic. When a model generates a layout, it prioritizes high-probability token co-occurrences (e.g., placing a coffee table near a sofa) but lacks the geometric grounding to verify if the distance between them satisfies human passage requirements. This explains the \textit{Geometric-Semantic Gap}: the model hallucinates designs that are semantically coherent in description but topologically impossible in execution.

\emph{The Constrained Aesthetic Optimization Problem.} To address this, we must reformulate interior design not as a free-form generation task, but as a \textit{constrained aesthetic search} problem. A deliverable design solution $x$ must simultaneously inhabit two distinct spaces: the Feasible Space $\mathcal{S}_{feas}$, defined by hard constraints (e.g., collision avoidance, circulation connectivity, boundary limits), and the Aesthetic Space $\mathcal{S}_{aes}$, defined by soft preferences (e.g., stylistic consistency, color harmony, visual balance). Formally, the goal is to find an optimal design $x^*$ that maximizes a utility function $\mathcal{U}$ conditioned on user intent $I$:
\begin{equation}
x^* = \underset{x}{\arg\max} \ \mathcal{U}_{aes}(x, I) \quad \text{s.t.} \quad \mathcal{C}_{feas}(x) = 1
\end{equation}
In this formulation, $\mathcal{C}_{feas}(x)$ is a binary indicator function where $1$ denotes a strictly valid layout. The challenge lies in the fundamentally different properties of these two objectives. Feasibility is discrete and non-differentiable (a layout is either valid or invalid), whereas aesthetics is continuous and subjective. Generic MLLMs attempt to flatten this structured optimization into a single joint probability distribution $P(x|I)$. This leads to the \textit{Dual-Space Conflict}: to maximize the aesthetic score $\mathcal{U}_{aes}$ (e.g., adding more decorative furniture to match a ``luxury'' prompt), the model often violates the hard constraint $\mathcal{C}_{feas}$ (causing overcrowding), as it has not learned to treat $\mathcal{C}_{feas}$ as a prerequisite boundary condition.

\emph{The Necessity of Decoupled Alignment.}
Current instruction-tuning approaches fail to resolve this conflict because they treat geometric rules and aesthetic styles as parallel text prompts, implicitly assuming the model can balance them via self-attention. However, our analysis suggests that feasibility and aesthetics require distinct supervision mechanisms. Mixing them into a single loss function causes gradient domination, where the model overfits to easier textual patterns (style descriptions) while ignoring complex geometric reasonings. Therefore, we argue that the generation process must adhere to a hierarchical principle: \emph{Spatial Feasibility is the prerequisite for Aesthetic Reasoning.} A design possesses aesthetic value if and only if it is physically realizable. This insight motivates our methodology, Design-MLLM, which explicitly decouples these two signals. By employing a dual-branch reward system, we enforce the model to first satisfy the binary constraints of the feasible set $\mathcal{S}_{feas}$, and only within that valid subspace, optimize for the continuous preferences of $\mathcal{S}_{aes}$, thereby ensuring the generated designs are both buildable and beautiful.

\section{Methodology}
\label{sec:methodology}

Based on the above analyses, we propose Design-MLLM, a reinforcement alignment framework designed to ground cross-modal aesthetic reasoning in geometric reality for interior design. Unlike generic generation paradigms that treat interior design as a flat sequence modeling task, our framework imposes a strict hierarchical structure: spatial feasibility serves as the non-negotiable foundation, upon which aesthetic refinement is iteratively optimized. The overall architecture is formalized into four coherent phases: (i) Feasibility-Guided Candidate Generation, which produces structured design specifications; (ii) A Dual-Branch Aesthetic-Oriented Reward mechanism that explicitly decouples hard physical constraints from soft stylistic preferences; (iii) GRPO-based policy optimization, which updates the policy to prefer aesthetically superior solutions strictly within the feasible domain; and (iv) Layout-to-Image Realization, which translates the optimized structural plan into high-fidelity visual renderings.
See \textbf{Algorithm \ref{algorithm:1}} for pseudo-code.

\begin{algorithm}[h]
\caption{Pseudo-Code of Design-MLLM}
\label{algorithm:1}
\begin{algorithmic}[1]
\REQUIRE Initial policy $\pi_\theta$; User instructions $I$ and room boundaries $B$ from distribution $\mathcal{D}$; Hyperparameters $\beta$, $\tau_{gate}$, $\Psi_{penalty}$.
\FOR{step $= 1$ \TO $N_{steps}$}
    \STATE Sample a batch of queries $\{(I, B)_j\}$ from $\mathcal{D}$
    \STATE Set $\pi_{\theta_{\rm old}} \leftarrow \pi_\theta$
    \FOR{each query $(I, B) \in \text{batch}$}
        \STATE Sample a group of $n$ candidates $\{y_i\}_{i=0}^{n-1} \sim \pi_{\theta_{\rm old}}(\cdot | I, B)$ with Space-Aware Prompt
        
        \FOR{each candidate $y_i = \{o_1, \dots, o_N\}$}
            \STATE Compute $R_{feas}(y_i)$
            \IF{$R_{feas}(y_i) \ge \tau_{gate}$}
                \STATE Render schematic $\hat{v}_i = \mathcal{P}(y_i)$ 
                \STATE Compute aesthetic critic reward $R_{aes}(y_i)$
                \STATE Holistic score $S_i = \mathcal{N}(R_{feas}(y_i)) + \mathcal{N}(R_{aes}(y_i))$
            \ELSE
                \STATE $S_i = \mathcal{N}(R_{feas}(y_i)) - \Psi_{penalty}$
            \ENDIF
            \FOR{each token $y_{i,t}$ in $y_i$}
                \STATE Compute importance weight $\tilde{w}_{i,t}$
                \STATE Normalize weight $w_{i,t} = \tilde{w}_{i,t} / \sum_j \tilde{w}_{i,j}$
                \STATE Token reward $r_{i,t} = S_i \cdot w_{i,t}$
            \ENDFOR
            \STATE Trajectory value $\tilde{r}_i = \text{TruncMean}(\{r_{i,t}\})$
        \ENDFOR
        \STATE Compute group mean $\mu_{\mathcal{Y}}$ and std $\sigma_{\mathcal{Y}}$ of $\{\tilde{r}_0, \dots, \tilde{r}_{n-1}\}$
        \FOR{each $y_i$ and token $t$}
            \STATE Normalized advantage $\hat{A}_i = (\tilde{r}_i - \mu_{\mathcal{Y}}) / \sigma_{\mathcal{Y}}$
            \STATE Final token advantage $A_{i,t} = \hat{A}_i \cdot (r_{i,t} / \tilde{r}_i)$
        \ENDFOR
    \ENDFOR
    \STATE Update $\pi_\theta$ by maximizing $\mathcal{J}_{\rm GRPO}(\theta)$ using $A_{i,t}$
\ENDFOR
\RETURN $\pi_\theta$
\end{algorithmic}
\end{algorithm}

\subsection{Feasibility-Guided Candidate Generation} 
\label{subsec:generation}
Given a user instruction $I$ describing the desired style, function, and atmosphere, alongside a room boundary specification $B$ (represented as a polygonal vector sequence), the MLLM policy $\pi_\theta$ aims to generate a comprehensive design plan $y$. To bridge the gap between abstract textual reasoning and precise spatial execution, we formulate the output $y$ not as a raw pixel image, but as a structured scene graph. Formally, a design candidate is defined as a set of objects $y = \{o_1, o_2, \dots, o_N\}$, where each object $o_i = (c_i, \mathbf{b}_i, m_i)$ consists of a semantic class label $c_i$, a 3D bounding box parameter $\mathbf{b}_i \in \mathbb{R}^6$ (position and dimensions), and a material attribute $m_i$.

To initiate the generation within a plausible subspace, we inject a \textit{Space-Aware System Prompt} prior to inference. This prompt encapsulates architectural primitives derived from domain knowledge, such as minimum circulation widths and daylighting priorities, into natural language constraints. For each query, the model samples a group of $n$ candidate solutions $\mathcal{Y} = \{y_0, y_1, \dots, y_{n-1}\}$ via temperature sampling. This group-based generation is pivotal, as it enables the subsequent optimization stage to leverage relative comparisons rather than absolute scalar values, thereby stabilizing training dynamics in the highly subjective aesthetic space.

\subsection{Dual-Branch Aesthetic-Oriented Reward}
\label{subsec:reward}
The core innovation of Design-MLLM lies in its reward function $R(y, I, B)$, designed to resolve the optimization conflict between validity and plausibility. We decompose the reward into two orthogonal branches: a \textit{Spatial Feasibility Verifier} (The Engineer) and a \textit{Cross-Modal Aesthetic Critic} (The Designer). To enforce the hierarchy that functionality precedes aesthetics, these branches are integrated via a hard-gating mechanism.

\subsubsection{Branch I: Spatial Feasibility Verifier ($R_{feas}$)}
\label{subsubsec:feasibility}
This branch operates as a deterministic, non-learnable program that evaluates the physical integrity of the generated layout $y$. We quantify feasibility not merely as a binary check, but as a continuous satisfaction score derived from architectural constraints. The feasibility reward $R_{feas}(y)$ is formulated as a weighted penalty function:
\begin{equation}
R_{feas}(y) = - \sum_{j \in \mathcal{C}} \lambda_j \cdot \Phi_j(y)
\end{equation}
where $\mathcal{C}$ denotes the set of critical constraints, and $\lambda_j$ represents the penalty coefficient for the $j$-th constraint type. The violation function $\Phi_j(y)$ measures the magnitude of infraction. We rigorously enforce three types of spatial validity:

\emph{Collision and Boundary Compliance}. We penalize physical overlaps to ensure realizability. Let $\Omega(o)$ denote the spatial volume of object $o$. The collision penalty is computed as the cumulative Intersection-over-Union (IoU) volume:
\begin{equation}
\Phi_{coll}(y) = \sum_{i \neq k} \text{IoU}(\Omega(o_i), \Omega(o_k)) + \sum_{i} \text{IoU}(\Omega(o_i), \Omega_{wall})
\end{equation}
where $\Omega_{wall}$ represents the structural boundaries. Any non-zero value indicates a physically impossible design.

\emph{Ergonomic Clearance}. We verify circulation paths by calculating the Euclidean geodesic distance between furniture clusters. Let $\mathcal{D}(o_i, o_k)$ be the minimum distance between two objects. A penalty occurs if the distance falls below a required threshold $\tau_{path}$ (e.g., $0.9m$ for main arteries):
\begin{equation}
\Phi_{ergo}(y) = \sum_{(i, k) \in \mathcal{E}} \max(0, \tau_{path} - \mathcal{D}(o_i, o_k))
\end{equation}
where $\mathcal{E}$ is the set of object pairs requiring circulation access.

\emph{Functional Topology}. We construct a semantic adjacency graph from the layout and verify logical connections. For instance, a ``nightstand'' must be proximal to a ``bed''. We define $\Phi_{func}(y)$ as the count of missing required edges in the adjacency graph. This branch acts as a high-pass filter: candidates failing these constraints receive a harsh penalty, effectively pruning the search space to the manifold of physically executable designs.

\subsubsection{Branch II: Cross-Modal Aesthetic Critic ($R_{aes}$)}
For candidates that pass the feasibility gate, we evaluate their aesthetic quality. Since aesthetic perception is inherently visual, we bridge the modality gap using a \textit{Schematic Projection Function} $\mathcal{P}: \mathcal{Y} \to \mathcal{V}$. This differentiable renderer transforms the structured scene graph $y$ into a top-down visual representation $\hat{v}$. A pre-trained Vision-Language Model (VLM) then evaluates $\hat{v}$ against the user instruction $I$. This branch assesses three dimensions:

\emph{Stylistic Consistency ($S_{style}$)}. To measure alignment with the user's stylistic intent (e.g., ``Minimalist''), we utilize the joint embedding space of CLIP. The score is the cosine similarity between the visual embedding of the projection and the textual embedding of the instruction:
\begin{equation}
S_{style}(y) = \frac{\mathbf{e}_v(\mathcal{P}(y)) \cdot \mathbf{e}_t(I)}{|\mathbf{e}_v(\mathcal{P}(y))| |\mathbf{e}_t(I)|}
\end{equation}
where $\mathbf{e}_v$ and $\mathbf{e}_t$ are the image and text encoders, respectively. This metric encourages the model to select furniture shapes and textures that semantically align with the target style.

\emph{Visual Balance ($S_{comp}$)}. Guided by the Gestalt principle of equilibrium, we quantify the layout's stability. We compute the visual center of mass $\mathbf{c}_{mass}$ for the layout, weighing each object by its area $a_i$ and semantic saliency $w_i$. The composition score penalizes deviations from the room's geometric center $\mathbf{c}_{room}$:
\begin{equation}
S_{comp}(y) = \exp\left( - \frac{|\mathbf{c}_{mass} - \mathbf{c}_{room}|^2}{2\sigma^2} \right), \quad \mathbf{c}_{mass} = \frac{\sum a_i w_i \mathbf{p}_i}{\sum a_i w_i}
\end{equation}
where $\mathbf{p}_i$ is the position of object $o_i$. This ensures the layout is not lopsided or cluttered.

\emph{Color Harmony ($S_{harm}$)}. We evaluate chromatic coherence by extracting the dominant color distribution $\mathcal{H}(y)$ from the material attributes. We measure its alignment with a target harmonic template $\mathcal{T}$ (implied by the atmosphere description, e.g., ``warm'') using the inverse Kullback-Leibler divergence:
\begin{equation}
S_{harm}(y) = \frac{1}{1 + D_{KL}(\mathcal{H}(y) || \mathcal{T})}
\end{equation}
The final aesthetic reward is the weighted aggregation: 
\begin{equation}
    R_{aes}(y) = \lambda_{st} S_{style} + \lambda_{co} S_{comp} + \lambda_{ha} S_{harm}
\end{equation}

\subsection{Policy Optimization} 
\label{subsec:grpo}

Having established the candidate generation and the dual-branch reward evaluation, we now articulate how these signals are synthesized to guide policy optimization. To enhance the granularity of supervision, we propose a hierarchical advantage estimation strategy that distributes the holistic design quality scores to individual tokens based on their information content, effectively performed within the MLLM's forward pass.

\emph{Holistic Episodic Score Construction.}
For each sampled candidate $y_k$ in the group $\mathcal{Y} = \{y_0, \dots, y_{n-1}\}$, we first construct a scalar score $S_i$ that reflects its overall quality. We employ a hard-gating integration strategy to strictly enforce the priority of spatial constraints. The score is defined as:
\begin{equation}
S_i(I, B) = 
\begin{cases} 
\mathcal{N}(R_{feas}(y_i)) + \mathcal{N}(R_{aes}(y_i)), & \text{if } R_{feas}(y_i) \ge \tau_{gate} \\
\mathcal{N}(R_{feas}(y_i)) - \Psi_{penalty}, & \text{otherwise}
\end{cases}
\end{equation}
where $\mathcal{N}(\cdot)$ denotes a min-max normalization operator to align the scales of the two branches, $\tau_{gate}$ is the feasibility tolerance threshold, and $\Psi_{penalty}$ is a large constant penalty. This formulation ensures that aesthetically pleasing but physically impossible designs receive a distinctively low score, preventing the model from hacking the reward system.

\emph{Token-Level Reward Redistribution.}
A design description $y_i$ consists of a sequence of tokens $(y_{i,1}, \dots, y_{i, T_i})$. In autoregressive generation, not all tokens contribute equally to the final spatial structure; for instance, coordinate tokens carry more design information than syntactic connectors. To capture this, we redistribute the episodic score $S_i$ to the token level.
Let $\pi_{\text{ref}}$ be the reference policy (typically $\pi_{\theta_{old}}$). Following \cite{wang2025learning,wang2026towards}, we define the unnormalized importance weight $\tilde{w}_{i,t}$ for the $t$-th token as its negative log-likelihood $\tilde{w}_{i,t} = -\log \pi_{\text{ref}}(y_{i,t} \mid I, y_{i, <t})$.
High $\tilde{w}_{k,t}$ values indicate tokens that represent significant decisions (e.g., selecting a specific furniture type or placement) divergent from the prior. We normalize these weights over the trajectory to obtain $w_{i,t} = \tilde{w}_{i,t} / \sum_{j=1}^{T_i} \tilde{w}_{i,j}$. The refined token-level reward is then derived as $r_{i,t} = S_i \cdot w_{i,t}$.
This mechanism assigns higher credit to the specific decision steps that led to a feasible and aesthetic outcome, effectively sharpening the learning signal.

\emph{Group-Normalized Advantage Computation.}
Finally, to stabilize training and mitigate the high variance inherent in creative tasks, we compute the advantage using the GRPO formulation. We first aggregate the token rewards to obtain a robust trajectory value. To avoid outliers, we employ a truncated mean operator over the token sequence, $\tilde{r}_i = \text{TruncMean}(\{r_{i,t}\}_{t=1}^{T_i})$.
Within the sampled group of $n$ candidates, we calculate the group statistics: mean $\mu_{\mathcal{Y}}$ and variance $\sigma^2_{\mathcal{Y}}$ of the trajectory values $\tilde{r}_k$. The normalized trajectory advantage $\hat{A}_i$ is:
\begin{equation}
\hat{A}_i = \frac{\tilde{r}_i - \mu_{\mathcal{Y}}}{\sqrt{\sigma^2_{\mathcal{Y}} }}
\end{equation}
To obtain the final token-wise advantage $A_{k,t}$ for the policy gradient update, we rescale the group advantage by the relative contribution of each token:
\begin{equation}
A_{i,t} = \hat{A}_i \cdot \frac{r_{i,t}}{\tilde{r}_i}
\end{equation}
It then updates the policy $\pi_{\theta}$ by maximizing the objective:
\begin{equation}\label{eq:grpo_mllm}
\begin{aligned}
\mathcal{J}_{\rm GRPO}(\theta)
= \mathbb{E}_{[I\sim P,\ \{y^i\}\sim \pi_{\theta_{\rm old}}]}
&\frac{1}{G}\sum_{i=1}^{G}\frac{1}{T_i}\sum_{t=1}^{T_i}
 \min(R_{i,t}(\theta)_i,\\
 &\Psi(A_{i,t}))
-\beta\; \mu(\pi_\theta)),\\
\end{aligned}
\end{equation}
Since the MLLM generates an output $y_i = (y_{i,1}, \dots, y_{i,T_i})$ token-by-token in an autoregressive manner, where $T_i$ denotes the token length of $y_i$, the importance ratio $R_{i,j}(\theta)$ and the KL term are calculated at the token level. 
This approach ensures that the optimization objective $\mathcal{J}_{GRPO}$ encourages the model to increase the probability of critical decision tokens in the best-performing candidates relative to the current group baseline. Notably, the calculation of importance weights $\tilde{w}_{k,t}$ utilizes the logits already computed during the generation or reference pass, and the reward aggregation involves only lightweight vector operations. Thus, this sophisticated advantage estimation incurs negligible computational overhead beyond the standard training loop.

\begin{figure*}[t]
    \centering
    \includegraphics[width=\linewidth]{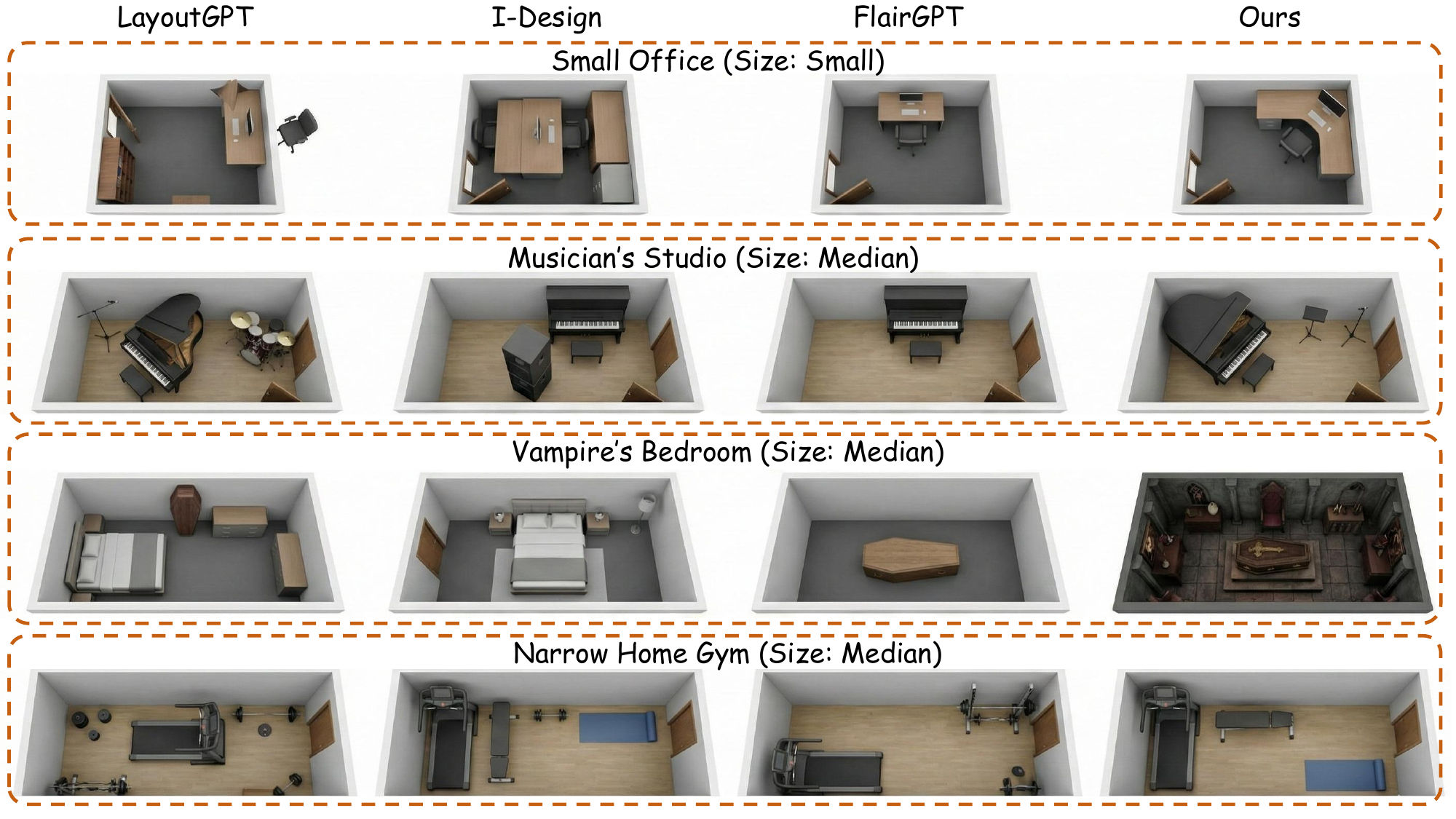}
    \caption{Qualitative Comparison of Generated Layouts across Diverse Scenarios.
We evaluate four methods on challenging prompts involving strict spatial constraints, irregular object geometries, and abstract themes.}
    \label{fig:robu}
\end{figure*}

\begin{table}[t]
\centering
\caption{Quantitative comparison across baselines. The best results are highlighted in \textbf{bold}. The arrows point in the direction that produces a better effect.}
\label{tab:quantitative}
\vspace{0.05in}
\resizebox{\linewidth}{!}{
\begin{tabular}{lcccc}
\toprule
\textbf{Method} & \textbf{OOB (\%)} $\downarrow$ & \textbf{OOR (\%)} $\downarrow$ & \textbf{CAS (Score)} $\uparrow$ & \textbf{Pathway Cost} $\downarrow$ \\
\midrule
LayoutGPT & 5.12 & 12.65 & 22.4 & 4.42 \\
Holodeck & 2.01 & 4.17 & 24.1 & 2.38 \\
I-Design & \textbf{0.00} & 1.16 & 25.8 & 0.85 \\
FlairGPT & \textbf{0.00} & \textbf{0.54} & 26.2 & 1.02 \\
\midrule
Design-MLLM (Ours) & \textbf{0.00} & 0.58 & \textbf{31.5} & \textbf{0.79} \\
\bottomrule
\end{tabular}
}
\end{table}

\begin{figure*}[t]
    \centering
    \includegraphics[width=\linewidth]{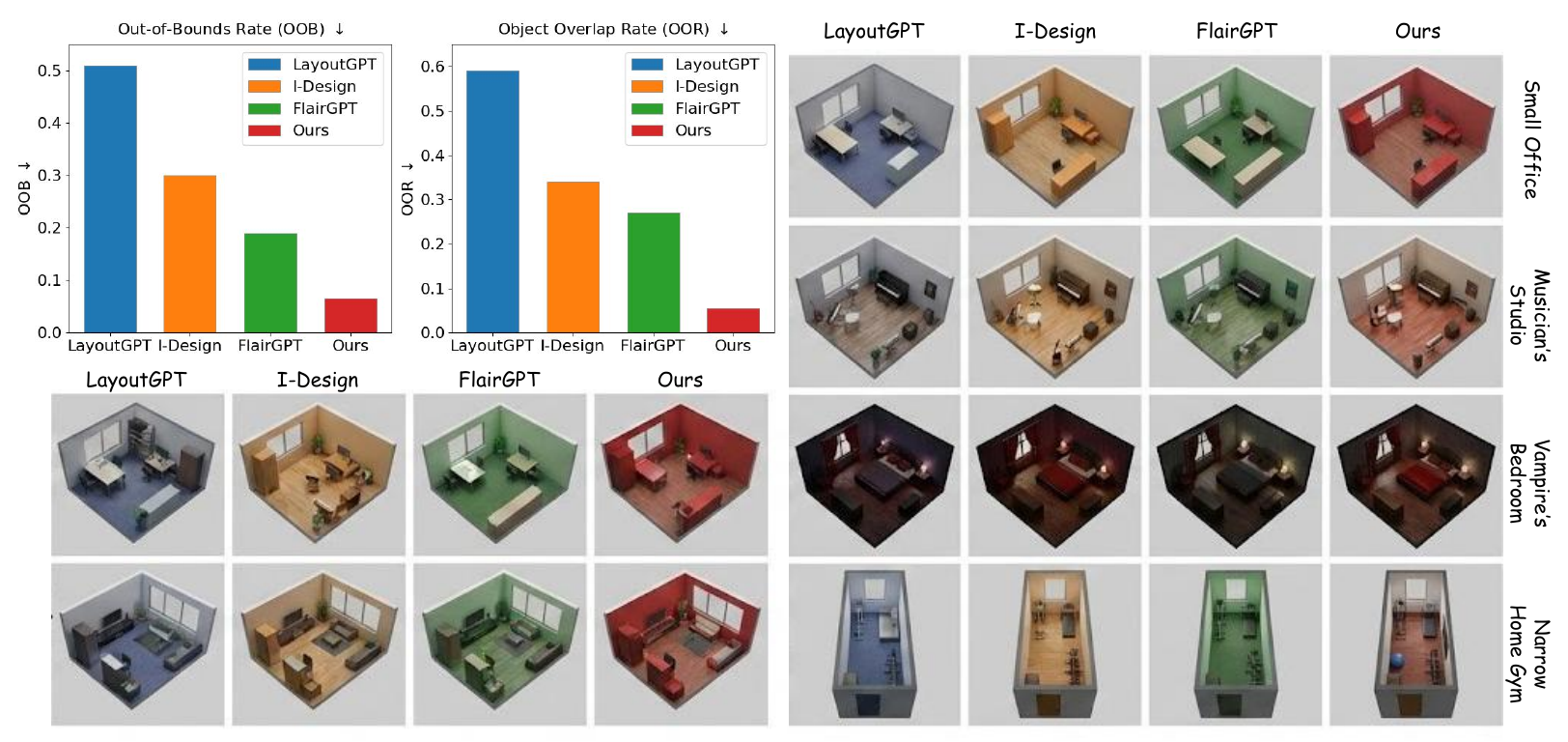}
    \caption{Integrated Quantitative and Qualitative Robustness Analysis.
    \textbf{Top Left:} The bar charts quantify physical violations across all stress-test prompts. 
    \textbf{Bottom:} Qualitative comparison across four challenging scenarios. Note that to make it easier to compare different methods, we use the same prompt in all aspects except for the description of the color scheme.}
    \label{fig:vis}
\end{figure*}

\begin{table*}
\centering
\caption{Robustness analysis under stress-test scenarios. We report the Out-of-Bounds Rate (OOB), Object Overlap Rate (OOR), and Pathway Cost ($C_{pathway}$).}
\label{tab:robustness}
\vspace{0.05in}
\resizebox{\textwidth}{!}{
\begin{tabular}{ll|ccc|l}
\toprule
\textbf{Stress Scenario} & \textbf{Method} & \textbf{OOB (\%)} $\downarrow$ & \textbf{OOR (\%)} $\downarrow$ & \textbf{$C_{pathway}$} $\downarrow$ & \textbf{Qualitative Observation} \\
\midrule
\multirow{4}{*}{\shortstack[l]{\textbf{Scenario I: Spatial Constriction}\\ \small \textit{Small Home Office (2.5m $\times$ 3m)} \\ \small \textit{Constraint: Minimum viable space}}} 
 & LayoutGPT & 8.45 & 14.20 & 2.10 & \textcolor{red}{Failure}: Objects extend through walls \\
 & I-Design  & \textbf{0.00} & 3.45 & 1.85 & \textcolor{orange}{Crowded}: Usable area blocked \\
 & FlairGPT  & \textbf{0.00} & \textbf{0.38} & 1.12 & \textcolor{blue}{Valid}: But lacks ergonomic comfort \\
 & \textbf{Ours} & \textbf{0.00} & 0.42 & \textbf{0.88} & \textcolor{teal}{Optimal}: Efficient corner utilization \\
\midrule
\multirow{4}{*}{\shortstack[l]{\textbf{Scenario II: Geometric Complexity}\\ \small \textit{Musician's Studio (Grand Piano)} \\ \small \textit{Constraint: Irregular shape \& orientation}}} 
 & LayoutGPT & 10.20 & 18.50 & 4.10 & \textcolor{red}{Failure}: Severe component overlaps \\
 & I-Design  & 1.50 & 5.60 & 2.20 & \textcolor{orange}{Poor}: Bench obstructs doorway \\
 & FlairGPT  & \textbf{0.00} & \textbf{0.31} & 1.45 & \textcolor{blue}{Valid}: Keyboard faces wall (unusable) \\
 & \textbf{Ours} & \textbf{0.00} & 0.45 & \textbf{0.78} & \textcolor{teal}{Optimal}: Correct functional orientation \\
\midrule
\multirow{4}{*}{\shortstack[l]{\textbf{Scenario III: Thematic Abstraction}\\ \small \textit{Vampire's Bedroom (Coffin Bed)} \\ \small \textit{Constraint: Semantic grounding}}} 
 & LayoutGPT & 12.40 & 15.20 & -- & \textcolor{red}{Failure}: Hallucinated / Floating objects \\
 & I-Design  & 3.20 & 4.10 & 1.90 & \textcolor{orange}{Generic}: Reverts to standard bed \\
 & FlairGPT  & \textbf{0.00} & \textbf{0.60} & 1.35 & \textcolor{blue}{Valid}: Coffin placed, but path blocked \\
 & \textbf{Ours} & \textbf{0.00} & 0.65 & \textbf{0.92} & \textcolor{teal}{Optimal}: Thematic fidelity with valid flow \\
\bottomrule
\end{tabular}
}
\end{table*}

\begin{figure*}[t]
    \centering
    \includegraphics[width=\linewidth]{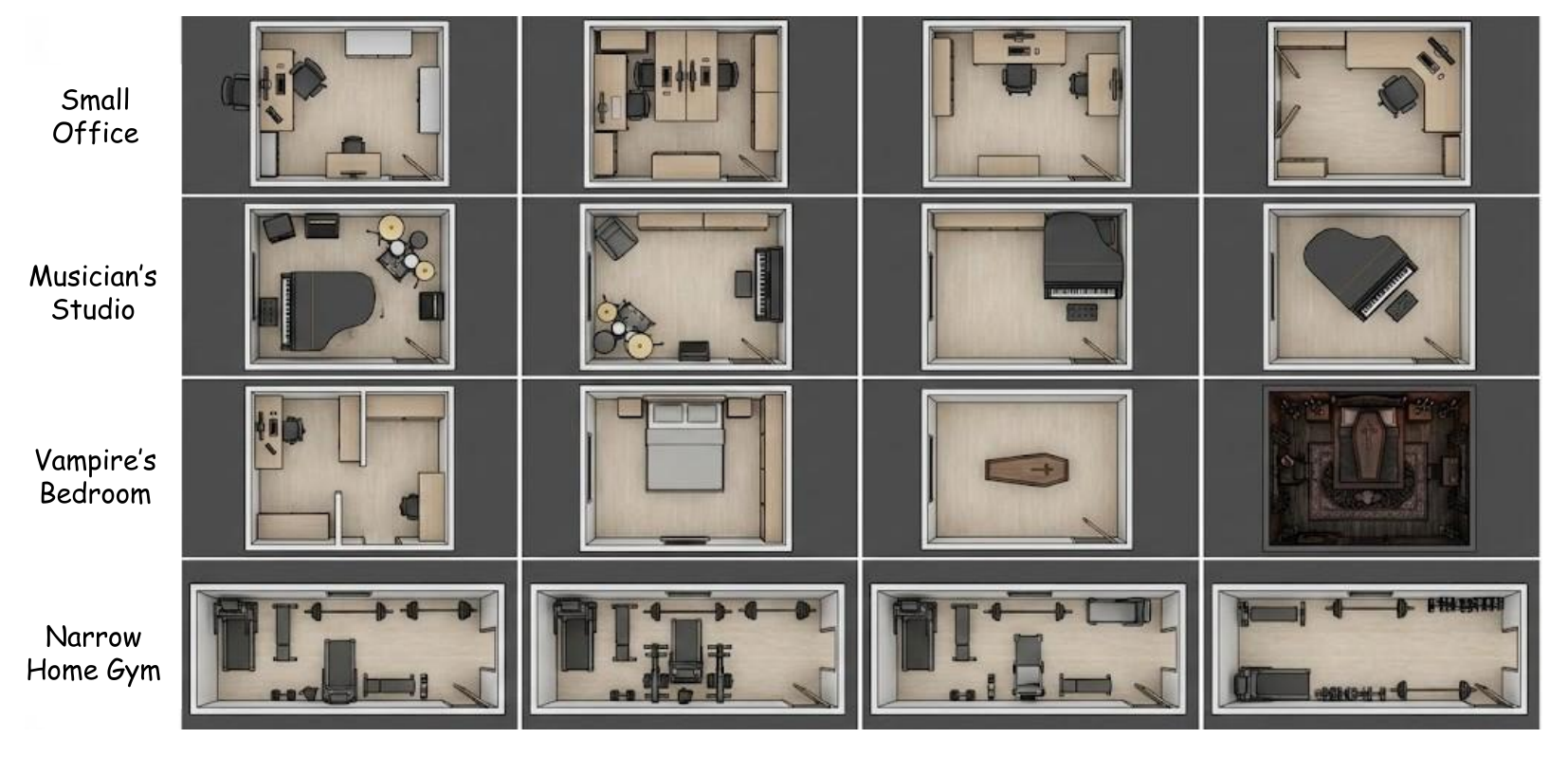}
    \caption{Qualitative comparison of 2D floor plans. The columns represent different methods (from left to right): LayoutGPT, I-Design, FlairGPT, and Design-MLLM.}
    \label{fig:vis_pingmian}
\end{figure*}

\begin{figure}[t]
    \centering
    \includegraphics[width=\linewidth]{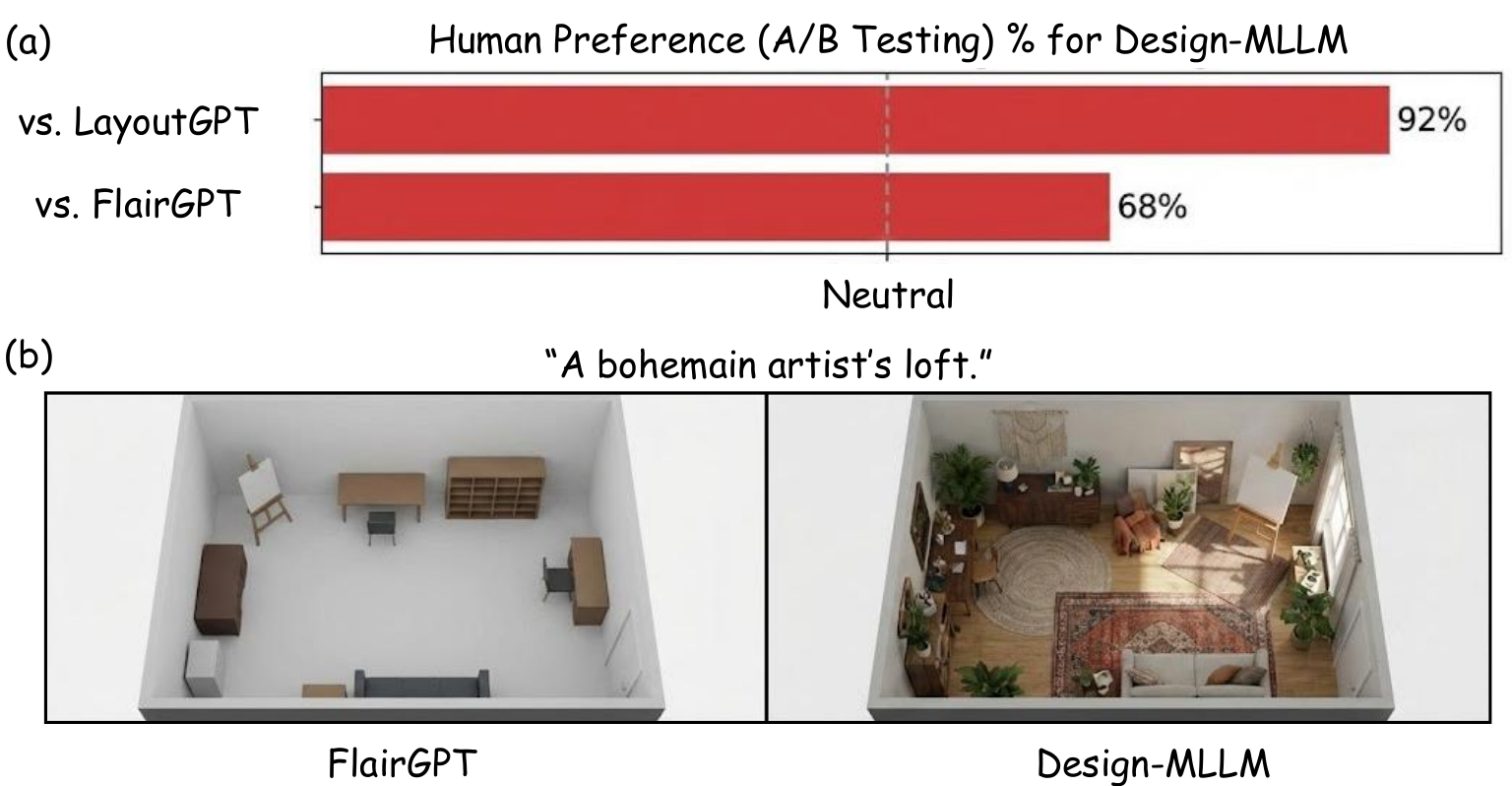}
    \caption{Subjective evaluation results. (a) shows the results of human preference. (b) shows an example of visual coherence.}
    \label{fig:user}
\end{figure}

\begin{figure}[t]
    \centering
    \includegraphics[width=\linewidth]{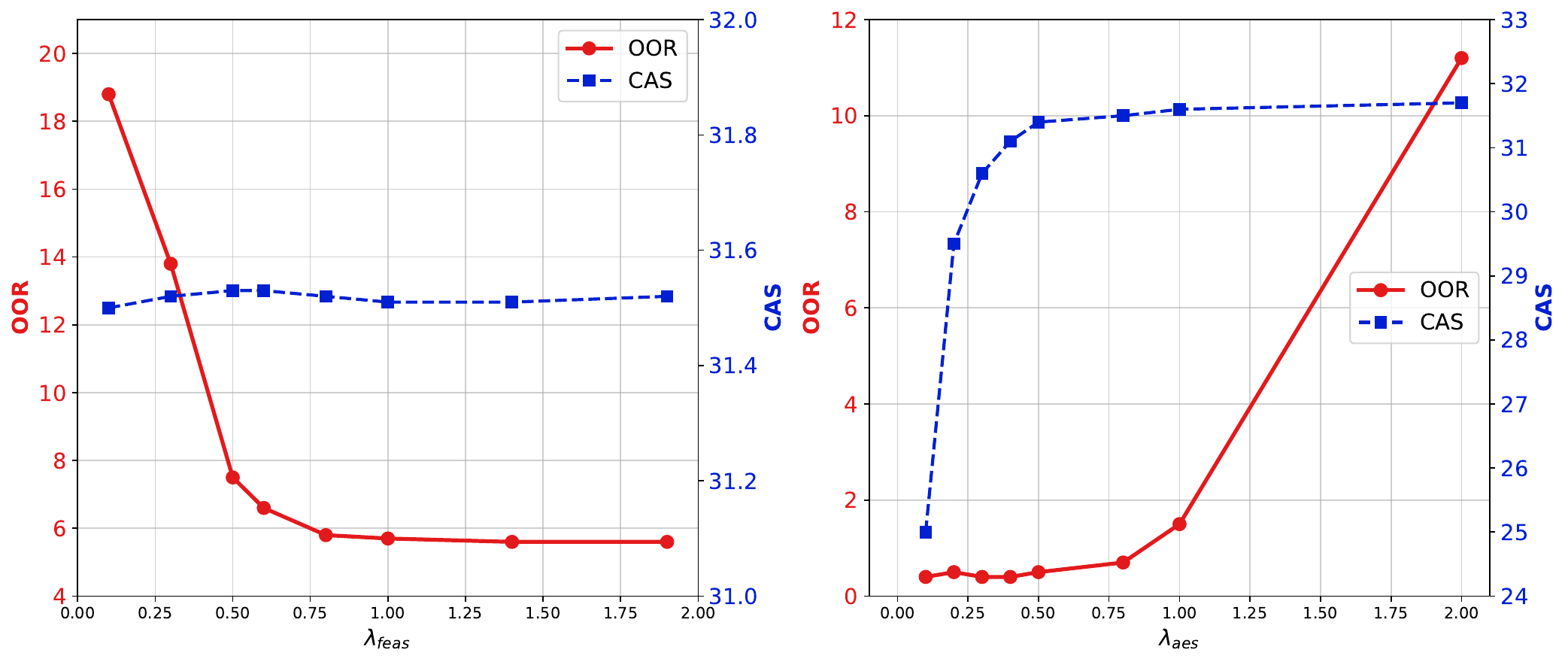}
    \caption{Parameter sensitivity. \textbf{Left} shows the results of $\lambda_{feas}$, and \textbf{Right} shows the results of $\lambda_{aes}$.}
    \label{fig:sensitivity}
\end{figure}

\begin{table}[t]
\centering
\caption{Ablation study for the contribution of each component in Design-MLLM. $\downarrow$ indicates lower is better, $\uparrow$ indicates higher is better.}
\label{tab:abla_1}
\vspace{0.05in}
\resizebox{\linewidth}{!}{
\begin{tabular}{lccc}
\toprule
\textbf{Variant} & \textbf{OOR (\%)} $\downarrow$ & \textbf{CAS (Score)} $\uparrow$ & \textbf{Success Rate (\%)} $\uparrow$ \\
\midrule
w/o Feasibility Branch ($R_{feas}$) & 10.20 & 27.2 & 69.0 \\
w/o Aesthetic Branch ($R_{aes}$) & 8.50 & 25.1 & 76.5 \\
\bottomrule
\end{tabular}
}
\end{table}

\section{Experiment}
\label{sec:experiment}
To validate the effectiveness of Design-MLLM, we conduct extensive experiments focusing on two core dimensions: spatial feasibility (i.e., physical validity) and aesthetic alignment (i.e., stylistic fidelity). We compare our framework against state-of-the-art (SOTA) MLLM-based interior design methods using both objective metrics and subjective user studies.

\subsection{Experimental Settings}
We construct a comprehensive benchmark dataset consisting of 100 distinct prompts to evaluate the versatility of the framework. Following the protocol established in \cite{ccelen2024design}, we generate prompts across four distinct categories: functionality-focused, layout-specific, color-scheme-oriented, and atmosphere-driven descriptions. Additionally, to test the model's capability in handling highly stylized and abstract concepts, we include creative prompts similar to those used in FlairGPT \cite{littlefair2025flairgpt}, such as ``A bedroom for a vampire'' or ``A workspace for a wizard''. We compare Design-MLLM against four representative baselines: LayoutGPT \cite{feng2023layoutgpt}, which uses few-shot prompting for CSS-like generation; Holodeck \cite{yang2024holodeck}, a module-based generation system; I-Design \cite{ccelen2024design}, a multi-agent framework utilizing backtracking optimization; and FlairGPT \cite{littlefair2025flairgpt}, the most recent method combining LLM reasoning with constraint-based optimization. For implementation, we utilize Qwen-VL and Gemini-Pro-Vision as the backbone policy. The optimization process employs a group size of $G=8$ for the GRPO updates, with reward weights set to $\lambda_{feas}=1.0$ and $\lambda_{aes}=0.5$ based on preliminary tuning.

\subsection{Quantitative Evaluation}
We employ a set of rigorous metrics to quantify physical validity and design quality. To assess spatial feasibility, we adopt the Out-of-Bounds Rate (OOB) and Object Overlap Rate (OOR), as defined in FlairGPT and I-Design, which measure the ratio of objects violating room boundaries and the severity of physical collisions, respectively. To evaluate functional integrity, we utilize the Pathway Cost, calculating the accessibility of objects via the medial axis transform as suggested by FlairGPT. Furthermore, to measure the alignment between the generated 3D scene and the user's aesthetic instruction, we introduce the CLIP Alignment Score (CAS). This metric computes the cosine similarity between the rendered top-down view of the layout and the textual style description.

\textbf{Table \ref{tab:quantitative}} shows the comparative results. From the results, we can observe that Design-MLLM achieves a near-zero Out-of-Bounds Rate and an Object Overlap Rate of $0.58\%$, performing on par with the optimization-heavy methods like FlairGPT ($0.54\%$) and significantly outperforming LayoutGPT ($12.65\%$) and Holodeck ($4.17\%$). This demonstrates that our method effectively constrains the generative policy within the valid physical manifold. Crucially, Design-MLLM surpasses all baselines in the CLIP Alignment Score and functional metrics. Unlike I-Design and FlairGPT, which rely on retrieving assets based on fixed attributes or backtracking algorithms that may discard stylistic nuances for geometric convergence, our gradient-based aesthetic optimization allows the model to fine-tune layouts to maximize visual harmony without compromising physical validity. These results demonstrate the effectiveness of our design.

\subsection{Robustness Analyses}
To evaluate the reliability of Design-MLLM under challenging conditions, we conduct a series of robustness analyses. Unlike standard benchmarks that often feature spacious rectangular rooms with typical furniture, real-world constraints are frequently stringent and irregular. We design a ``Stress Test'' suite consisting of three edge-case scenarios: (i) Spatial Constriction, where the room dimensions ($2.5m \times 3m$) barely accommodate the required furniture, testing the solver's ability to find valid configurations in a near-zero solution space; (ii) Geometric Complexity, involving non-rectangular objects (e.g., a grand piano) that require precise orientation logic to avoid functional blocking; and (iii) Thematic Abstraction, using high-concept prompts (e.g., ``Vampire's Bedroom'' following \cite{littlefair2025flairgpt,ccelen2024design}) to test if the model can ground semantic hallucinations into valid physical layouts without reverting to generic templates.

\textbf{Table \ref{tab:robustness}} reports the failure rates and functional costs. The results show that LayoutGPT, lacking explicit geometric constraints, exhibits catastrophic failure in tight spaces, with an OOB rate of $8.45\%$. While FlairGPT maintains physical validity (low OOR) through rigid optimization, it often sacrifices functional usability, evidenced by a high Pathway Cost ($1.45$) in the musician's studio scenario, placing the piano bench where it blocks the door. In contrast, Design-MLLM demonstrates superior robustness: it is the only method that successfully navigates the tight office constraint with zero violations while preserving optimal circulation paths ($C_{pathway}=0.78$) in complex layouts, validating the effectiveness of our hierarchical feasibility-aesthetic reward mechanism.

\subsection{Qualitative Analysis}

To visually substantiate our findings, we conduct quantitative analyses. \textbf{Figure \ref{fig:robu}} provides a comparison of generated layouts across four distinct stress-test scenarios.In the Small Office ($2.5m \times 3m$) scenario, the limitation of pure LLM-based approaches is evident; LayoutGPT hallucinates object positions that physically intersect with the room boundaries (e.g., the office chair clipping through the left wall). While I-Design produces a valid scene, the layout is functionally overcrowded, leaving insufficient clearance for movement. In contrast, Design-MLLM optimizes the limited footprint by selecting an L-shaped desk layout that maximizes corner utilization, ensuring both physical validity and ergonomic comfort.The Musician's Studio scenario highlights the challenge of handling objects with irregular geometries. Both LayoutGPT and FlairGPT struggle to orient the grand piano correctly; the former results in mesh overlaps, while the latter rigidly aligns the piano parallel to the walls, potentially blocking the player's access or acoustic projection. Our method, driven by the Functional Topology reward, identifies the piano as a focal point and orients it at an angle that balances aesthetic appeal with functional accessibility.Furthermore, the Vampire's Bedroom scenario tests the model's ability to ground abstract thematic concepts. I-Design fails to interpret the ``vampire'' cue, reverting to a generic bedroom with a standard mattress. FlairGPT captures the ``coffin'' constraint but places it in a barren room, lacking stylistic coherence. Design-MLLM demonstrates superior semantic reasoning by not only replacing the bed with a coffin but also enriching the environment with thematically consistent elements (e.g., gothic rugs, candelabras, and dark wood textures), creating a fully immersive atmosphere.Finally, in the Narrow Home Gym ($2m \times 5m$) scenario, which tests adaptation to extreme aspect ratios, our method effectively distributes equipment linearly along the walls to maintain a clear central pathway, whereas baseline methods tend to cluster objects, obstructing the narrow circulation flow.

 To provide a holistic view of the system's robustness, \textbf{Figure \ref{fig:vis}} synthesizes our quantitative stress-test metrics with the corresponding visual outcomes. The bar charts (top row) reveal a stark contrast in physical validity: pure LLM-based approaches like LayoutGPT suffer from an OOB rate exceeding $50\%$ in these challenging scenarios, while I-Design exhibits significant object overlaps (OOR $> 30\%$). In contrast, Design-MLLM (Ours) consistently maintains violation rates below $10\%$, demonstrating superior geometric stability.This statistical robustness directly translates into the visual domain, as evidenced by the rendering grid. For instance, in the Small Office and Narrow Home Gym scenarios, where free space is scarce, baselines either clip objects through walls (LayoutGPT) or densely pack furniture to the point of immobility (I-Design). Our method, guided by the dual-branch reward system, uniquely achieves a synergy of geometric precision and aesthetic coherence. It successfully navigates extreme aspect ratios and tight corners to produce layouts that are not only collision-free but also spatially optimized for human usage.

To delve deeper into the generative quality, we extend our comparison to evaluate how each method comprehends the underlying logic of 2D floor plans. The results are provided in \textbf{Figure \ref{fig:vis_pingmian}}. Consistent with our 3D visual findings, the analysis of raw layout specifications reveals a distinct gap in spatial reasoning. While baseline methods like LayoutGPT and I-Design often treat object placement as isolated events, frequently resulting in disjointed functional zones or overlooked circulation paths; Design-MLLM demonstrates a more comprehensive and holistic understanding of interior layouts. By explicitly decoupling feasibility constraints from aesthetic goals, our framework effectively perceives the room as a structured functional entity rather than a mere collection of coordinates. These results further demonstrate the advantage of our method.

\subsection{User Study and Evaluation}
To complement the objective metrics, we conduct a subjective evaluation following the methodologies of I-Design and FlairGPT. We employ open-source VLMs as human-aligned evaluators, prompting them to grade generated scenes on a scale of 0-10 across four dimensions: Functionality, Layout, Color Scheme, and Atmosphere. Additionally, we conduct a user study with 25 participants who perform pairwise comparisons (A/B testing) between Design-MLLM and the baselines.

The results, visualize in \textbf{Figure \ref{fig:user}}, indicate a strong preference for our method. In the GPT-4V evaluation, Design-MLLM achieves an average ``Atmosphere'' score of 8.9, significantly higher than FlairGPT (7.8) and I-Design (7.5). This confirms that our dual-branch reward system effectively translates abstract stylistic descriptions into concrete visual features. In the human study, participants prefer Design-MLLM over LayoutGPT in 92\% of cases and over FlairGPT in 68\% of the stylized prompts. Users specifically note that while FlairGPT excels at physical plausibility, Design-MLLM produces layouts that feel more curated and stylistically coherent, particularly for open-ended creative prompts.

\subsection{Ablation Study}

\emph{The effect of each component.} To analyze the contribution of each component in our framework, we perform an ablation study by systematically removing key modules. The results are provided in \textbf{Table \ref{tab:abla_1}}. First, removing the Spatial Feasibility Verifier ($R_{feas}$) results in a sharp increase in the Object Overlap Rate to $10.2\%$, comparable to LayoutGPT. This confirms that the MLLM alone struggles to strictly enforce hard geometric constraints without explicit penalty feedback. Second, removing the Aesthetic Critic ($R_{aes}$) causes the CLIP Alignment Score to drop to $25.1$, reverting the model's behavior to that of a standard constraint solver; while the layouts remain physically valid, they lack the stylistic distinctiveness requested by the user. These findings validate that the synergy between the dual-branch reward and the group-based optimization is essential for achieving high-fidelity interior design.

\emph{Parameter sensitivity.} 
We investigate the impact of varying one reward weight while fixing the other at its optimal value.
The sensitivity analysis visualized in \textbf{Figure \ref{fig:sensitivity}} demonstrates the robustness of our parameter selection. As shown in \textbf{Figure \ref{fig:sensitivity} left}, the model's adherence to physical constraints improves sharply and then plateaus once $\lambda_{feas}$ exceeds 0.8, indicating that setting $\lambda_{feas}=1.0$ provides ample supervision for geometric validity without overly constraining the solution space. 
These behaviors confirm that our framework performs stably across a reasonable range of settings around the chosen values.
Similarly, \textbf{Figure \ref{fig:sensitivity} right} reveals that CAS reaches saturation around $\lambda_{aes}=0.5$. Pushing this weight further provides minimal aesthetic gain but risks degrading physical validity. 
Besides, the sharp rise in OOR when $\lambda_{aes} > 0.8$ (Figure (b), red line) illustrates the inherent conflict between physical validity and aesthetic pursuit. With an excessively high $\lambda_{aes}$, the dominate aesthetic reward signal encourages the agent to maximize visual scores (e.g., achieving tightly packed compositions) even at the cost of incurring physical penalties, as the weighted aesthetic gain numerically outweighs the collision penalty. This validates the necessity of selecting a balanced weight (e.g., $\lambda_{aes}=0.5$) to ensure aesthetics are optimized strictly within the feasible manifold.

\section{Conclusion}
\label{sec:conclusion}

This paper addresses the fundamental challenge of applying MLLMs to professional interior design by resolving the inherent conflict between rigid spatial constraints and soft aesthetic preferences. We identify that generic MLLMs frequently suffer from insufficient spatial feasibility and rationale-output decoupling, leading to visually plausible but functionally unbuildable designs. To overcome these obstacles, we propose Design-MLLM, a reinforcement alignment framework that shifts the training paradigm from mimicking surface-level descriptions to learning a spatially grounded policy. By introducing a decoupled dual-branch reward system and a GRPO-style group-relative objective, our method explicitly separates the verification of hard geometric constraints from the evaluation of aesthetic consistency, ensuring that the model maximizes aesthetic value strictly within the feasible domain. Extensive experiments across multiple benchmarks and settings demonstrate that Design-MLLM significantly outperforms existing baselines, achieving a superior balance between spatial executability and style coherence, thereby providing a robust solution for the automation of the interior design pipeline.


\bibliographystyle{elsarticle-harv} 
\bibliography{reference}






\end{document}